\documentclass[%
 reprint,
superscriptaddress,
amsmath,
amssymb,
aps,
pra,
longbibliography
]{revtex4-2}

\usepackage[version=3]{mhchem} % Formula subscripts using \ce{}
\usepackage{lipsum}

\usepackage{longtable}% FOR SUPPLEMENT

\usepackage{placeins}
\usepackage{xspace}
\usepackage{appendix}
\usepackage[detect-all]{siunitx}
\usepackage[normalem]{ulem}
\usepackage{multirow}
\usepackage{booktabs}
\usepackage{colortbl}

\usepackage{array}

\newcolumntype{C}[1]{>{\centering\let\newline\\\arraybackslash\hspace{0pt}}m{#1}}

\usepackage{graphicx}% Include figure files
\usepackage{dcolumn}% Align table columns on decimal point
\usepackage{bm}% bold math
\usepackage{booktabs}
\usepackage{hyperref}% add hypertext capabilities
\usepackage{xcolor}
\hypersetup{
    colorlinks,
    linkcolor={blue!70!black},
    citecolor={blue!70!black},
    urlcolor={blue!70!black}
}
\usepackage{dcolumn}
\newcolumntype{L}{D{.}{.}{2,5}}
\newcommand*{\wn}{\ensuremath{\text{cm}^{-1}}\xspace}

\newcommand{\T}{$\mathcal{T}$}

\begin{document}

\title{Hyperfine-Resolved Spectroscopy of Dysprosium Monoxide (DyO) for Precision Measurements of the Nuclear Schiff Moment}
\author{Zack D. Lasner}
\altaffiliation{Current address: IonQ, Inc., College Park, MD 20740, USA}
\affiliation{Harvard-MIT Center for Ultracold Atoms, Cambridge, Massachusetts 02138, USA}
\affiliation{Department of Physics, Harvard University, Cambridge, Massachusetts 02138, USA}
\author{Aidan T. Ohl}
\affiliation{Department of Chemistry, Williams College, Williamstown, MA 01267, USA}
\author{Nicole M. Albright}
\affiliation{Department of Chemistry, Williams College, Williamstown, MA 01267, USA}
\author{Kendall L. Rice}
\affiliation{Department of Chemistry, Williams College, Williamstown, MA 01267, USA}
\author{Charlene Peng}
\affiliation{Department of Chemistry, Williams College, Williamstown, MA 01267, USA}
\author{Lan Cheng}
\affiliation{Department of Chemistry, The Johns Hopkins University, Baltimore, MD 21205, USA}
\author{John M. Doyle}
\affiliation{Harvard-MIT Center for Ultracold Atoms, Cambridge, Massachusetts 02138, USA}
\affiliation{Department of Physics, Harvard University, Cambridge, Massachusetts 02138, USA}
\author{Benjamin L. Augenbraun}
\email{bla1@williams.edu}
\affiliation{Department of Chemistry, Williams College, Williamstown, MA 01267, USA}

\date{November 14, 2025}

\begin{abstract}
\noindent We perform laser spectroscopy of dysprosium monoxide (DyO) to determine the hyperfine structure of the ground X8 and excited [17.1]7 states in the $^{161}$Dy and $^{163}$Dy isotopologues. These dysprosium nuclei have non-zero nuclear spin and dynamical octupole deformation, which gives them high sensitivity to time-reversal-violating new physics via the nuclear Schiff moment (NSM). The DyO molecule was recently identified as being amenable to optical cycling---the basis for many laser cooling and quantum control techniques---which makes it a practical candidate for NSM searches. The measurements reported here are prerequisites to implementing optical cycling, designing precision measurement protocols, and benchmarking calculations of molecular sensitivity to symmetry-violating effects. The measured hyperfine parameters are interpreted using simple molecular orbital diagrams and show excellent agreement with relativistic quantum chemical calculations. 
\end{abstract}

\maketitle

\section{Introduction}

Overwhelming evidence has accumulated over recent decades that the Standard Model of particle physics is incomplete~\cite{Young2016survey}. Measurements of effects that violate time-reversal (\T) symmetry in polarized atoms and molecules, including the nuclear Schiff moment (NSM) of heavy nuclei, have been among the most fruitful avenues to probe new physics at energy scales of $\gtrsim$10~TeV~\cite{ACME2014,AndreevACME,cairncross2017precision,JilaEDM,Graner2016reduced,Cho1989tenfold,Cho1991search,safronova_search_2018}. NSM measurements in atoms are limited by incomplete polarization of the atomic orbitals, but in molecules the effective orbital polarization---and corresponding sensitivity to the NSM---can reach three orders of magnitude higher. Moreover, the NSM of a high-$Z$ nucleus possessing an octupole deformation and nonzero nuclear spin can offer another three-orders-of-magnitude enhanced sensitivity to symmetry-violating interactions~\cite{sushkov1948possibility, flambaum2020electric, haxton1983enhanced}. In addition to the statically deformed nuclides of Fr, Ra, Ac, Th, and Pa~\cite{dalton2023enhanced}, certain lanthanide nuclei can exhibit a dynamical, vibration-like octupole deformation that leads to a sizable NSM~\cite{spevak1997enhanced}. 
For example, octupole deformation in Dy is well established theoretically~\cite{rodriguez2023beyond, dalton2023enhanced}.

The ultimate sensitivity of an NSM measurement scales with the number of particles probed and the coherence time, in addition to the intrinsic nuclear sensitivity and degree of orbital polarization. These factors favor experiments with many long-lived, cold, trapped molecules. The key requirement for direct cooling, as well as the quantum-state manipulation required for many high-fidelity measurement techniques, is the existence of an optical cycle~\cite{mccarron2018laser, fitch2021lasercooled}. A recent theoretical study identified $^{161}$DyO as a promising candidate for NSM searches~\cite{chen_relativistic_2024}. In addition to the intrinsic sensitivity of the $^{161}$Dy nucleus, it was posited that DyO could be cooled to ultracold temperatures and trapped using optical cycling. Finally, the molecular sensitivity parameter of DyO was computed to be on the same order as other important platforms for future NSM measurements including ThF$^+$, FrAg, and RaOH. Because dysprosium metal is inexpensive, abundant, and non-radioactive, DyO may enable an experimentally straightforward route to prepare large numbers of molecules in traps with long coherence times. In a similar manner, europium-containing solids offer an intriguing pathway to high-sensitivity NSM measurements~\cite{Ramachandran2023nuclear}.

The lanthanide monoxides have long been of interest in physical chemistry because they offer tests of ligand field theory models as metal cations $M^{2+}$ perturbed by the field of oxygen anions O$^{2-}$~\cite{field1982diatomic, liu_laser_1984, linton_analysis_1988}. The low-lying electronic states of DyO were characterized as far back as 1957~\cite{gatterer1957molecular}. Pioneering work by the Linton group analyzed the rotational structure of DyO in the ground X8 and excited [17.1]7 electronic states~\cite{linton_laser_1986}. The Linton group also performed experiments at improved resolution that were capable of resolving the hyperfine structure of the $[17.1]7 \leftarrow X8$ (0,1) vibronic transition, as well as the rotational structure of the [18.0]9 excited state~\cite{cheng_laser_1993}. The electric dipole moment of DyO has also been measured by the Linton group~\cite{linton_stark_1992}. All of these studies demonstrated that DyO could be well described by simple principles of ligand-field theory, providing intuitive descriptions of the electronic structure. More recently, there has been renewed interest in DyO and DyO$^+$ seeking to characterize its Rydberg series, ionization energy, and bond dissociation energy~\cite{schaller_gas-phase_2024, schaller_cold_2024}.

Here, we report high-resolution laser excitation spectroscopy of the [17.1]7$\leftarrow$X8 band to measure the hyperfine structure of $^{161}$DyO and $^{163}$DyO, including the previously unresolved electric quadrupole interaction. These measurements are required for the construction of closed optical cycling transitions in future molecular cooling and trapping experiments, as well as the design of concrete precision measurement protocols. Our results also provide insight into the short-range electronic wave functions, showing very good agreement with theoretical values.

\section{Experimental Methods}
Molecular beams of DyO were produced via the reaction of laser-ablated dysprosium with nitrous oxide (\ce{N2O}). A 6.3~mm diameter rod of Dy (99.7\% purity) was ablated by the second harmonic of a pulsed Nd:YAG laser operating at 10~Hz with an ablation energy of approximately 20~mJ. Continuous rotation and translation of the Dy rod ensured a fresh surface for each ablation pulse. A mixture of approximately 5\% \ce{N2O} in argon was introduced through a pulsed valve at a relatively high backing pressure of 3000~kPa to produce rotational cooling to $T \lesssim 10$~K. The opening of the pulsed valve was timed to entrain the ablation products before expanding through a nozzle into a vacuum chamber maintained at a typical running pressure of $8\times 10^{-5}$~Torr.

To resolve the rotational and hyperfine structure of the [17.1]7$\leftarrow$X8 transition, the molecular beam was probed 50~cm downstream from the source after passing through a 1~cm diameter collimating aperture. The molecules were excited by the output of a continuous-wave dye laser (linewidth $\sim$1~MHz) operating with Rhodamine 6G dye. We recorded excitation spectra on both the [17.1]7$\leftarrow$X8 (0,0) and (0,1) vibronic bands near 586~nm and 616~nm, respectively. Laser-induced fluorescence (LIF) at 586~nm was collected by an in-vacuum lens system and focused onto a photomultiplier tube (PMT). In a typical experiment, the laser was stabilized to a specific laser frequency using a commercial wavemeter and 10 ablation pulses were averaged together. The resulting signal was integrated over the $\sim$200~$\mu$s temporal width of the molecular beam. The laser frequency was then stepped by 10 MHz and the process repeated in order to generate an excitation spectrum.

\subsection{Observations}

\begin{figure*}[tb]
    \centering 
    \includegraphics[width=0.85\linewidth]{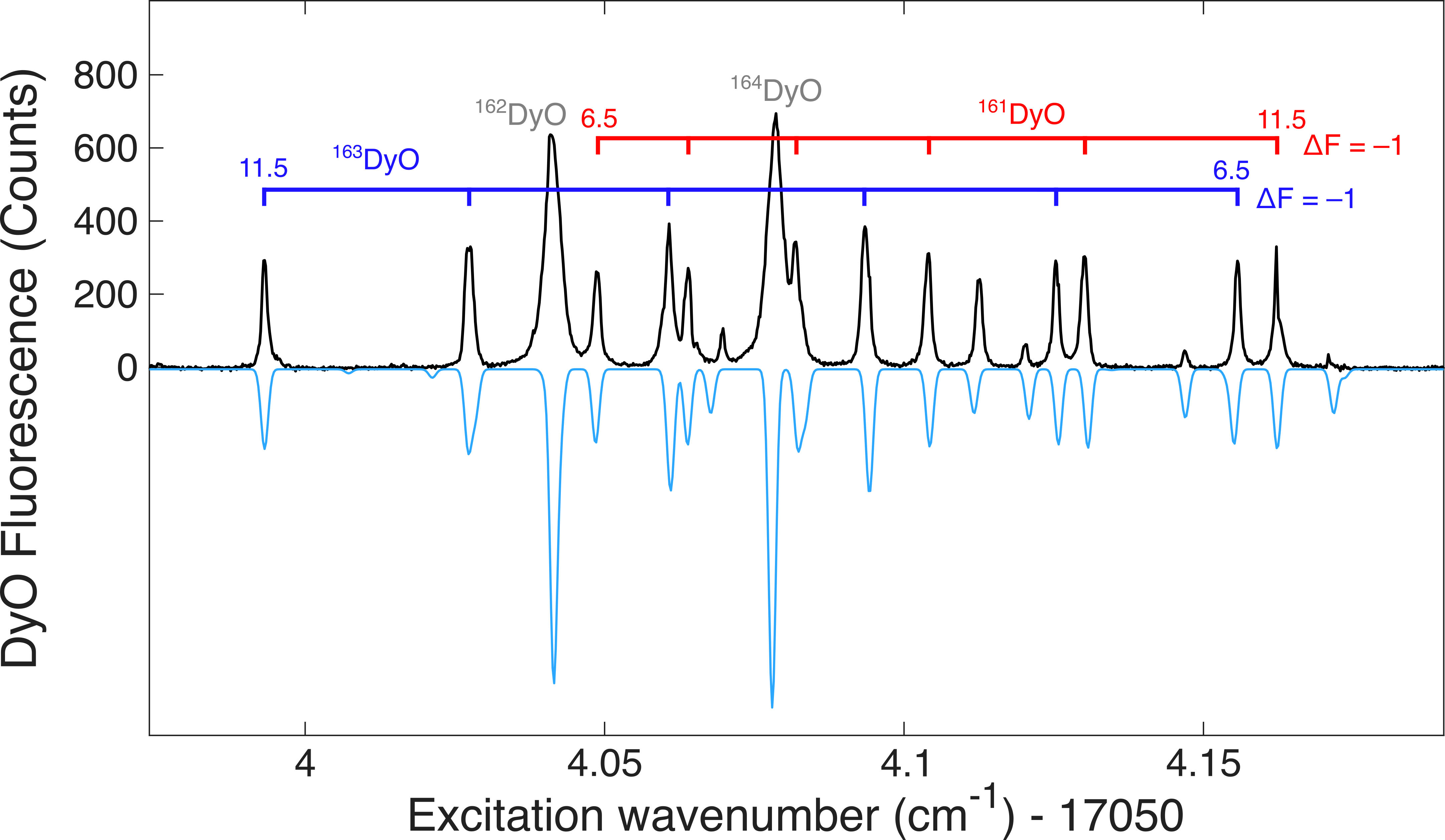}
    \caption{Observed (black, upward-going) and simulated (blue, downward-going) spectrum of the [17.1]7$\leftarrow$X8 (0,0) P(9) branch features. Horizontal lines connect the sets of main ($\Delta F = -1$) peaks for the P branch. The individual hyperfine components are labeled by their values of $F$.}
    \label{fig:DyOOriginBand}
\end{figure*}

We originally recorded several rotational features of the [17.1]7$\leftarrow$X8 (0,0) vibronic band in order to investigate the hyperfine structure in the X8 and [17.1]7 states. Each rotational branch feature was highly congested due to the presence of four abundant isotopologues ($^{161-164}$DyO), where both $^{161}$DyO and $^{163}$DyO have nuclear spin $I_\text{Dy} = 5/2$. This means that each rotational branch feature has at least 14 strong $\Delta F = \Delta J$ hyperfine components within a typical span of 0.2~\wn. As an example, the P(9) branch feature is shown in Figure~\ref{fig:DyOOriginBand}.  While the excitation features are all well resolved, it was difficult to pick out hyperfine sequences and to assign peaks to each isotopologue.

\begin{figure*}[tb]
    \centering 
    \includegraphics[width=0.9\linewidth]{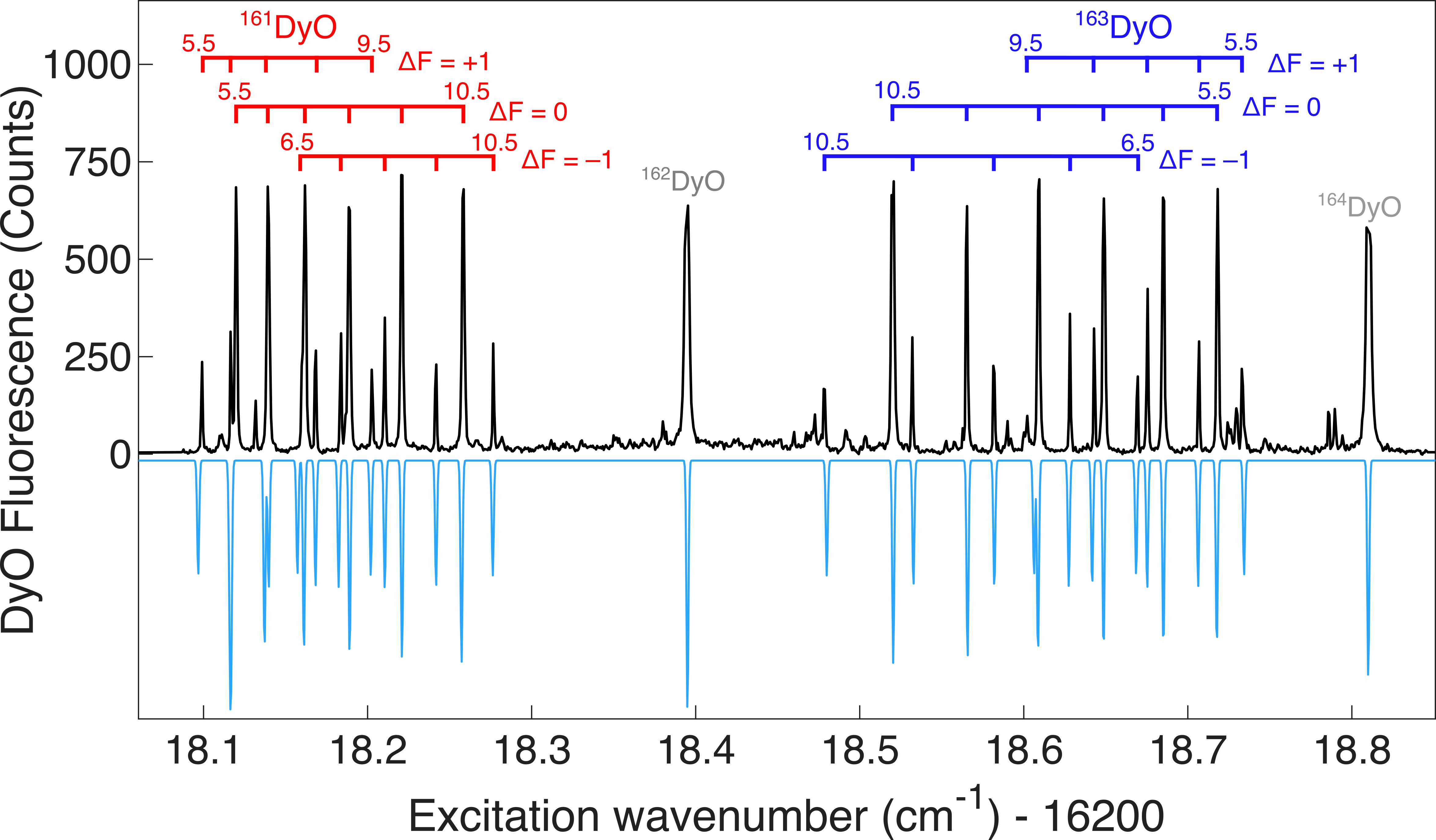}
    \caption{Observed (black, upward-going) and simulated (blue, downward-going) spectrum of the [17.1]7$\leftarrow$X8 (0,1) Q(8) branch features recorded at high laser power. Horizontal lines connect the sets of main and satellite transitions, which have $\Delta F = 0$ and $\Delta F=\pm1$, respectively, for the Q branch. The individual hyperfine components are labeled by their values of $F$.}
    \label{fig:DyOHotBand}
\end{figure*}

To separate the signals originating from each isotopologue, we recorded portions of the [17.1]7$\leftarrow$X8 (0,1) band. The Q(8) branch feature of this band, plotted in Figure~\ref{fig:DyOHotBand}, demonstrates that the isotope shifts in the [17.1]7$\leftarrow$X8 (0,1) band are sufficiently large to provide unambiguous assignment of peaks to each isotopologue. Because the vibrational excitation was in the ground state of the transition, transitions for lighter isotopologues fell at lower frequencies than those for the heavier isotopologues. 

At low laser power (1 mW), only the main ($\Delta F = \Delta J$) hyperfine components were observed, and these could be assigned based on their relative intensities. We also recorded portions of the spectrum at higher laser power (80 mW; the condition shown in Figure~\ref{fig:DyOHotBand}), which allowed us to observe satellite ($\Delta F \neq \Delta J$) transitions. The observation of satellite peaks is beneficial because these transitions provide direct hyperfine combination differences that can be used to confirm the quantum number assignments.

After assigning the (0,1) band, we returned to the (0,0) band. Because the hyperfine parameters are only weakly dependent on the vibrational state, the primary difference between each rotational branch feature in the (0,0) and (0,1) bands is the relative isotope shift. Thus, it was relatively straightforward to assign features in the (0,0) band by looking for hyperfine combination differences within the ground-state energy levels that matched those from the (0,1) band. 

Prior work by the Linton group, reported in the Ph.D. thesis of C.-H. Cheng~\cite{cheng_laser_1993}, had used combination differences to estimate the magnetic hyperfine constant, $h$, in the [17.1]7$\leftarrow$X8 (0,1) vibronic band. The combination differences determined in the present work agree well with those from Ref.~\onlinecite{cheng_laser_1993}. However, because our analysis achieved higher resolution, we are able to determine both $h$ and $eQq_0$ for the X8($v=0$), X8($v=1$), and [17.1]7$(v'=0)$ states in a global fit.

\subsection{Analysis}
We fit the spectrum to a conventional effective Hamiltonian model of the rotational and hyperfine structure. For the least-squares fit, the energy levels of each state were represented by the effective Hamiltonian
\begin{equation}
    H = T_0 + H_\text{r} + H_\text{m} + H_\text{q},
\end{equation}
representing the contributions of the state origin, rotational, magnetic hyperfine, and electric quadrupole hyperfine terms, respectively. All matrix elements are taken from Ref.~\onlinecite{dick2015hyperfine}. We use a Hund's case (c) basis set due to the large spin-orbit coupling in DyO. Matrix elements of the rotational Hamiltonian are
\begin{equation}
    \langle J\Omega \lvert H_\text{r} \rvert J\Omega\rangle= B[ J(J+1) - \Omega^2] - D [ J(J+1)-\Omega^2 ]^2.
\end{equation}
Though we have only recorded the lowest few rotational branch features, we included the centrifugal distortion constant for the excited state to help account for the strong perturbations that were  observed in previous rotationally-resolved studies of DyO~\cite{linton_laser_1986}.

For the magnetic hyperfine interaction, the determinable parameter in Hund's case (c) coupling is $h$, which is a linear combination of the Frosch and Foley hyperfine parameters~\cite{frosch1952magnetic}, $h = a \Lambda + \left(b_F + \frac{2}{3}c \right)\Sigma$. The diagonal matrix elements are:
\begin{align}
    \langle J\Omega F \lvert H_\text{m} \lvert J \Omega F \rangle= h \Omega \, \frac{F(F+1)-J(J+1)-I(I+1)}{2J(J+1)}.
\end{align}
Second-order perturbation theory estimates indicated that off-diagonal hyperfine terms contribute energy-level shifts smaller than our measurement uncertainty and were thus excluded. Finally, the electric quadrupole hyperfine interaction ($eQq_0$) is represented for our purposes by the diagonal term only, with matrix elements
\begin{align}
     \langle J\Omega F \lvert H_\text{q} \lvert J \Omega F \rangle  = eQq_0 \, f(I,J,F) \left( \frac{3\Omega^2}{J(J+1)} -1 \right),
\end{align}
where
\begin{equation}
f(I,J,F) = \frac{0.75C(C+1)-I(I+1)J(J+1)}{2I(2I-1)(2J-1)(2J+3)}
\end{equation}
and
\begin{equation}
    C = F(F+1)-J(J+1)-I(I+1).
\end{equation}

We performed global least-squares fits to a total of 195 observed lines for the (0,0) and (0,1) bands of $^{161}$DyO and $^{163}$DyO. A list of the observed and calculated transition frequencies is available in the Supporting Information (Tables S1--S4)~\cite{SupportingInfo}. The fits reproduced the observed data to within a standard deviation of 0.0021~\wn (0.0016~\wn) for $^{161}$DyO ($^{163}$DyO), which is commensurate with the typical measurement uncertainty. The resulting rotational and hyperfine parameters for $^{161}$DyO and $^{163}$DyO are reported in Table~\ref{tab:RotationalFit}. 

\begin{table*}[tb]
\setlength{\tabcolsep}{5pt}
		\caption{Rotational and hyperfine parameters for the the X8$(v=0,1)$ and [17.1]7$(v'=0)$ states of $^{161}$DyO and $^{163}$DyO. All values are reported in \wn. Error bars represent the 1$\sigma$ uncertainties.}
		\label{tab:RotationalFit}
        \setlength\tabcolsep{0pt}
\begin{tabular*}{0.9\linewidth}{@{\extracolsep{\fill}}cccccccc}
\hline \hline
        & \multicolumn{3}{c}{$^{161}$DyO}                  &  & \multicolumn{3}{c}{$^{163}$DyO}                 \\ \cline{2-4} \cline{6-8} 
        & X8($v=0$) & X8($v=1$)       & {[}17.1{]}7($v'=0$) &  & X8($v=0$) & X8($v=1$)      & {[}17.1{]}7($v'=0$) \\ \hline
$T_v$   & 0         & $842.3890(96)$ &  $17057.1551(11)$ &  & 0         & $841.92373(71)$ &        $17057.11029(80)$            \\
$B$     & $0.359014(45)$   & $0.357528(35)$      & $0.266297(90)$         &  & $0.358644(32)$ & $0.357174(27)$     & $0.266221(64)$         \\
$D$     &       &              & $-0.000301(2)$       &  &          &              & $-0.000299(1)$       \\
$h$     & $-0.05109(49)$  & $-0.05116(44) $     & $-0.02437(49)$         &  & $0.07143(34)$   & $0.07138(32)$      & $0.03408(34)$          \\
$eQq_0$ & $0.0096(35)$ & $0.0162(28)$        & $0.0347(37)  $         &  & $0.0098(25) $       & $0.0101(20)$       & $0.0349(26)$           \\ \hline \hline
\end{tabular*}
\end{table*}

Simulated spectra were generated by taking differences between calculated energy levels using the parameters output by our least-squares fitting. The relative intensity, $S$, of each component was computed via the formula~\cite{manke_electronic_2008}
\begin{equation}
    S(J'F' \leftarrow JF) \propto (2F+1)(2F'+1) \begin{Bmatrix}
        J' & F' & I \\ F & J & 1
    \end{Bmatrix}^2
\end{equation}
and a Gaussian lineshape of width 30 MHz was applied to all simulated peaks. To qualitatively account for saturation effects, a logarithmic scaling was applied to the simulated intensity axis for comparison to datasets recorded at high power.

\section{Discussion}

\subsection{Validity of the molecular parameters}
Several consistency checks can be applied to the molecular parameters reported in Table \ref{tab:RotationalFit}. The ground-state rotational constants are very similar to the value of $B$ found for $^{164}$DyO in Ref.~\onlinecite{linton_laser_1986}. As was pointed out in Ref.~\onlinecite{linton_laser_1986}, the [17.1]7 state displays significant perturbations. Since we have only measured transitions from a few of the lowest $J$ states, our excited-state $B$ value is most appropriately treated as an effective parameter that characterizes the rotational energy level spacing of the excited state. The value of $B$ for the [17.1]7 state ($\sim$0.266~\wn) appears to be in good agreement with the effective value of the rotational constant shown in Figure 3 of Ref.~\onlinecite{linton_laser_1986}.  Moreover, the isotopic scaling of the rotational constant is $B^{(163)}/B^{(161)} = 1.0013$, which is close to the expected value of $(\mu_{163}/\mu_{161}) = 1.0011$. The energy separation between X8($v=0$) and X8($v=1$) provides an estimate of the vibrational constant of X8. The ratio of isotopic values $\omega_e^{(163)}/\omega_e^{(161)} \approx 1.00055$ compares very favorably to the theoretical reduced-mass scaling of $(\mu_{163}/\mu_{161})^{1/2} = 1.000549$. Similar checks can be applied to the magnetic hyperfine constants. The ratio of the magnetic hyperfine parameter $h$ for $^{161}$DyO and $^{163}$DyO is $h^{(163)}/h^{(161)} \approx -1.40$ for all three states studied in this work. This ratio is very close to the ratio of nuclear $g$-factors, $g_N^{(163)}/g_N^{(161)} = -1.3995$, providing confidence in the determined values. Similarly, the fitted values of $eQq_0$ are consistent between $^{161}$DyO and $^{163}$DyO for all three states, as expected for $Q^{(163)}/Q^{(161)} \approx 1.06$ and relative uncertainties in fitted $eQq_0$ values exceeding 6\%.

\subsection{Interpretation of the Hyperfine Constants}

\begin{figure}[tb]
    \centering 
    \includegraphics[width=0.95\linewidth]{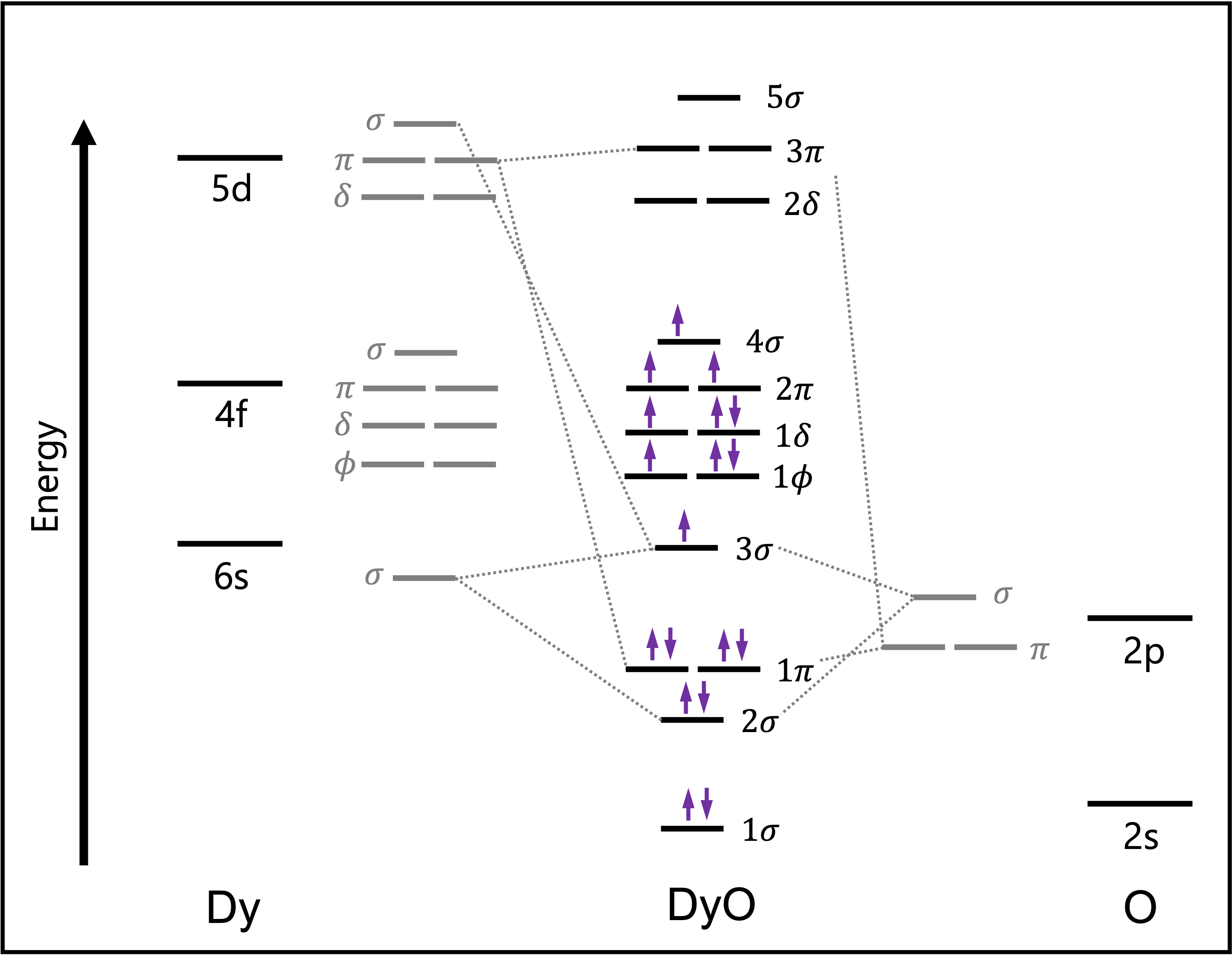}
    \caption{Qualitative correlation diagram showing the ground-state electron configuration. Atomic energy levels for Dy and O are shown on the outside, ligand-field splittings are shown in gray in the intermediate region, and the resulting molecular orbital energy levels of DyO are given in the center.}
    \label{fig:MOdiagram}
\end{figure}

Ligand field theory successfully describes the essential features of the energy level structure of lanthanide monoxides. The ground-state configuration of DyO is dominated by the Dy$^{2+}$[$4f^9 6s^1$]O$^{2-}$ atomic configuration. The lowest-energy state of the Dy $4f^9$ core is the $^{6}H_{7.5}$ state, i.e. $S_c=2.5, L_c=5$ with aligned spin $S_c$ and orbital angular momentum $L_c$ to generate the total angular momentum of the core electrons, $J_c = 7.5$. Coupling $J_c$ to the unpaired $6s$ electron leads to a pair of states with total atomic angular momentum $J_a = 7$ and $8$. The $J_a = 8$ state is lower in energy due to the exchange interaction. Each $J_a$ state of the Dy$^{2+}$ moiety has allowed (signed) angular momentum projections on the internuclear axis given by $\Omega = \pm J_a, \pm(J_a-1), \ldots, 0$. States with $\Omega=\pm|\Omega|$ are degenerate and considered to lie in the same electronic manifold, while those with different values of $|\Omega|$ are split by the electric field arising from the O$^{2-}$ ligand. The lowest-energy state has $|\Omega| = J_a$; thus the ground state of DyO has $|\Omega| = 8$ and is labeled as X8. Since the X8 state originates predominantly from an atomic configuration Dy$^{2+}$[$4f^9(^{6}H_{7.5}) 6s^1(^7H_8)$]O$^{2-}$, the electronic configuration of X8 is approximately described by the Slater determinant
\begin{align}\label{eq:slater-determinant}
\Psi(X_8) \approx 
\lvert & s\sigma_0^\alpha,\, f\phi_{-3}^\alpha,\, f\phi_{+3}^\alpha,\, f\phi_{+3}^\beta, \nonumber\\
& f\delta_{-2}^\alpha,\, f\delta_{+2}^\alpha,\, f\delta_{+2}^\beta,\, 
f\pi_{-1}^\alpha,\, f\pi_{+1}^\alpha,\, f\sigma_0^\alpha \rangle,
\end{align}
where each subscript and superscript denotes the orbital angular momentum and spin projection, respectively, of a single-electron state. A molecular orbital diagram, shown with this configuration occupied in Figure \ref{fig:MOdiagram}, can be useful to rationalize the behavior of the low-lying states.

\subsubsection{Magnetic Hyperfine Interaction}

Following the method detailed in Ref.~\onlinecite{dick2015hyperfine}, the dominant contributions to $h$ can be estimated in a semi-empirical manner from tabulated values of atomic wave function integrals. The determinable parameter $h$ is a combination of the Frosch and Foley hyperfine parameters~\cite{dick2015hyperfine, frosch1952magnetic}, $h = a \, \Lambda + \left(b_F + \frac{2}{3}c \right) \Sigma$. The ground-state configuration of DyO arises from \ce{Dy^{2+}} $4f^96s^1$ and the term symbol is well described by ${}^7H_{8}$ ($\Lambda = 5$ and $\Sigma = 3$). This means that for the ground state of DyO, 
\begin{equation}
    h = 5a + 3 b_F + 2 c.
    \label{eq:habc}
\end{equation}

Each of these parameters can be related to an expectation value of the electronic wave function (with all expressions in units of Hz)~\cite{dick2015hyperfine}:
\begin{align}
    a &=  K \left(\frac{1}{\Lambda}\right) \left\langle \Lambda \left\lvert \sum_{i} \frac{\ell_{zi}}{r_i^3} \right\rvert \Lambda \right\rangle \label{eq:a} \\
    b_F &= \left( \frac{8\pi}{3} \right) K \left( \frac{1}{S}  \right) \left\langle \Lambda, \Sigma \left\lvert \sum_{i} s_{zi} \delta_i(r) \right\rvert \Lambda, \Sigma \right\rangle  \label{eq:bF} \\
    c&= \frac{3}{2} K \left(\frac{1}{S}\right)  \left\langle \Lambda, \Sigma \left\lvert \sum_{i} s_{zi} \frac{(3 \cos^2 \theta_i - 1)}{r_i^3} \right\rvert \Lambda, \Sigma \right\rangle. \label{eq:c}
\end{align}
Here, $K = \left( \frac{\mu_0}{4\pi h_P} \right) g_e g_N \mu_B \mu_N$, where $\mu_0$ is the permeability of free space, $h_P$ is Planck's constant, $g_e$ is the (positive) electron $g$-factor, $g_N$ is the nuclear $g$-factor, $\mu_B$ is the Bohr magneton, $\mu_N$ is the nuclear magneton, the sums run over all electrons, $\hat{z}$ is along the internuclear axis, and the coordinate origin is the Dy nucleus.

Numerical evaluation of Equations~\ref{eq:a}--\ref{eq:c} requires computing integrals over atomic wave functions. We take the simple approach of substituting values obtained from Hermann-Skillman wave functions~\cite{morton_atomic_1978}, $r_i^{-3} \rightarrow \langle r_i^{-3}\rangle=10.59\,a_0^{-3}$ and $\langle \delta_i(r)\rangle \rightarrow \lvert \psi(0) \rvert^2 = 6.459\,a_0^{-3}$, where $a_0$ is the Bohr radius and $\psi(0)$ is the electronic wave function evaluated at the Dy nucleus (non-vanishing only for the $6s$ electron).  We note that provided these substitutions are approximately valid, $a$ is independent of the orbital configuration, while $b$ depends only on the number of unpaired $s$-shell electrons and total spin. The value of $c$ depends on the relative concentration of the wave functions for electrons with unpaired spin around the equator of the Dy atom as compared to the poles (producing smaller or larger values of $\langle 3\cos^2\theta-1 \rangle$, respectively), in addition to the total spin. Therefore $c$ is highly sensitive to the polarization of the electron cloud, which is not captured by the simple model of the atomic configuration considered here. Nevertheless, we employ Eq.~\ref{eq:slater-determinant} for the molecular orbital configuration in order to estimate the order-of-magnitude scale of $c$.

Using the \textit{ab initio} value of $\langle r^{-3} \rangle$~\cite{morton_atomic_1978}, with experimental nuclear $g$-factors $g_N^{(163)} = 0.269$ and $g_N^{(161)} = -0.192$~\cite{ferch_hyperfine_1974}, we can evaluate numerical values of $h$. We find $a^{(161)} = -0.00648$~\wn, $b_F^{(161)} = -0.00552$~\wn, $c^{(161)} = -0.00108$~\wn, and $a^{(163)} = 0.00907$~\wn, $b_F^{(163)} = 0.00772$~\wn, and $c^{(163)} = 0.00151$~\wn. We note that $|2c|\ll|5a+3b_F|$ for both isotopes, so that the estimated value of $h$ is relatively insensitive to the details of molecular polarization. Overall, we predict $h^{(161)} = -0.0511~\text{ cm}^{-1}$ and $h^{(163)} = 0.0715~\text{ cm}^{-1}$ for the electronic ground state. These values compare very well against the experimental measurements of $h^{(161)} = -0.0511(5)$~\wn and $h^{(163)} = 0.0714(3)$~\wn (see Table~\ref{tab:RotationalFit}).

Alternatively, we can use experimental measurements of $^{161}$Dy hyperfine structure to provide the effective radial integral parameterized as $\langle r^{-3} \rangle^{01} \approx 8.64$ a.u.~\cite{childs_hyperfine_1984, buttgenbach_hyperfine_1982}. This value, which comes from the leading magnetic hyperfine term, is approximately 20\% smaller than the theoretically computed value. In this case, we obtain $h^{(161)} = -0.0447~\text{ cm}^{-1}$ and $h^{(163)} = 0.0626~\text{ cm}^{-1}$. The good agreement with the experimental values in Table~\ref{tab:RotationalFit} suggests that the ground state of DyO is well modeled by the configuration of Eq. \ref{eq:slater-determinant}.

\subsubsection{Electric Quadrupole Interaction}
The electric quadrupole coupling constant, $eQq_0$, can be estimated via~\cite{dick2015hyperfine}
\begin{equation}\label{eq:eQq0}
    eQq_0 = K' \left \langle \Lambda,\Sigma=S \left\lvert \sum_i \frac{3\cos^2\theta_i-1}{r_i^{3}} \right\rvert \Lambda,\Sigma=S \right\rangle.
\end{equation}
Here, $K' = \left(\frac{-e^2}{4\pi\epsilon_0} \right) Q$, where $\epsilon_0$ is the permittivity of free space and $Q$ is the nuclear quadrupole moment, with $Q^{(161)}=2.51$~barns and $Q^{(163)}=2.65$~barns~\cite{stone_table_2013, tanaka_systematics_1984}. By using atomic integrals similar to those involved in calculating the $c$ hyperfine parameter, we can estimate $eQq_0$. For the case of the electric quadrupole interaction, we use the experimentally-derived value of $\langle r^{-3} \rangle^{02} = 6.96\,a_0^{-3}$ obtained from the leading electric quadrupole interaction ($b^{02}$) in atomic Dy~\cite{childs_hyperfine_1984}. We obtain $eQq_0^{(161)} \approx 0.091~\text{ cm}^{-1}$ and $eQq_0^{(163)} \approx 0.096~\text{ cm}^{-1}$. 

Overall, it is satisfying that a very simple molecular orbital analysis predicts small, positive values for $eQq_0$ that are in reasonable agreement with the measured values. Nonetheless, the predicted values are approximately an order of magnitude larger than the measured values of $eQq_0$, which are all approximately $0.01~\text{ cm}^{-1}$. To explain this discrepancy, we recall the fact that $\langle3\cos^2\theta-1\rangle$ in Eq.~\ref{eq:eQq0} probes the polarization of the electronic wave function along the Dy--O axis. This polarization is highly sensitive to the precise admixtures of different atomic states in the molecular orbitals, and the simple model used here underestimates the tendency of the valence electrons to concentrate along the poles around the Dy nucleus. To illustrate likely state mixing in Fig.~\ref{fig:MOdiagram} beyond the simple configuration of Eq.~\ref{eq:slater-determinant}, we include dotted lines from the Dy 5d states into the 1$\pi$ and 3$\sigma$ molecular orbitals, as well as mixing of O orbitals into 2$\sigma$ and 3$\sigma$. It is also expected that mixing with higher-lying orbitals (such as Dy~6p$\sigma$) and polarization of the core electrons could further reduce the magnitude of $eQq_0$.

\subsection{\emph{Ab initio} calculations}

We have used the CFOUR program package~\cite{CFOURfull,matthews2020coupledcluster} to perform relativistic quantum chemistry calculations focusing on the properties of the electronic ground $X8$ state of DyO.
The wave function of the $X8$ state is dominated by a Dy($6s^1 5f^{9}$)O$^{2-}$ configuration. 
In the spinor representation, i.e., with spin-orbit coupling included in the orbitals,
the leading determinant for the ground state wave function with $\Omega=8$ can be written as
\begin{align}\label{eq:spinordeterminant}
\lvert (\text{core}) ~ (4f_{\frac{5}{2}})^6  ~ 6s_{\frac{1}{2},\frac{1}{2}} ~ 4f_{\frac{7}{2},\frac{7}{2}} ~ 4f_{\frac{7}{2},\frac{5}{2}} ~ 4f_{\frac{7}{2},\frac{3}{2}} \rangle,
\end{align}
and that with $\Omega=-8$ as
\begin{align}\label{eq:spinordeterminant2}
\lvert (\text{core}) ~ (4f_{\frac{5}{2}})^6  ~ 6s_{\frac{1}{2},-\frac{1}{2}} ~ 4f_{\frac{7}{2},-\frac{7}{2}} ~ 4f_{\frac{7}{2},-\frac{5}{2}} ~ 4f_{\frac{7}{2},-\frac{3}{2}} \rangle.
\end{align}
Here we use a notation $nl_{j,m_j}$ to represent a molecular orbital dominated by
an atomic spinor with $n$ as the principal quantum number, $l$ being the spectroscopic symbol for the orbital angular momentum,
$j$ as the total angular momentum, and $m_j$ being the projection of the total angular momentum on the internuclear axis.  
In Eqs. \ref{eq:spinordeterminant} and \ref{eq:spinordeterminant2}, 
``core'' represents the fully occupied Dy 1s, 2s, 2p, 3s, 3p, 3d, 4s, 4p, 4d, 5s, 5p orbitals and O 1s, 2s, 2p orbitals,
and ``$(4f_{\frac{5}{2}})^6$'' denotes the fully occupied Dy $4f_{\frac{5}{2}}$ subshell.

We have first performed relativistic two-component Kramers unrestricted Hartree-Fock (HF) calculations
for the electron configuration in Eq. \ref{eq:spinordeterminant}
and then treated dynamic electron correlation using 
the coupled-cluster singles and doubles (CCSD)~\cite{Purvis82} augmented with a non-iterative triples [CCSD(T)]~\cite{Raghavachari89} method. The exact two-component Hamiltonian~\cite{Dyall97,Kutzelnigg05,Ilias07,Liu09} augmented with atomic mean-field~\cite{Hess96a} spin-orbit integrals, an X2CAMF scheme based on the Dirac-Coulomb-Gaunt Hamiltonian~\cite{Liu18,Zhang22}, has been used 
to treat relativistic and spin-orbit coupling effects. 
The uncontracted ANO-RCC basis set for Dy~\cite{Faegri01,Roos08} 
and the uncontracted aug-cc-pVTZ basis set for O~\cite{Kendall92} have been used 
in all calculations presented here. 
Forty-eight core electrons including Dy 1s, 2s, 2p, 3s, 3p, 3d, 4s, 4p, 4d electrons and O 1s electrons 
as well as virtual spinors with orbital energies higher than 100 Hartree
have been kept frozen in the CC calculations.

To determine the equilibrium bond length and force constants,
we have calculated the X2CAMF-CCSD(T) energies in 11 grid points 
with bond lengths $r=1.800 + 0.02 n, n=0, \pm1, \pm2, \pm3, \pm4, \pm5$ and fitted
the potential energy curve as a sixth-order polynomial of the bond length.
The anharmonic constants have been obtained using the computed force constants and second-order vibrational perturbation theory (VPT2)~\cite{Mills72}.
The calculations of the electric dipole moments, 
magnetic hyperfine constants and Dy electric quadrupole-coupling constants have used
the recent implementation of analytic X2CAMF-CCSD(T) gradients~\cite{liu21}.
These X2CAMF-CC calculations have benefited from the efficient implementation
using atomic-orbital based algorithms~\cite{Liu18b}. 
The Gaussian nuclear charge distributions \cite{Visscher97} have been used for all calculations presented here.

The X2CAMF-CCSD(T) equilibrium bond length $r_e$ takes a value of 
1.793 \AA. This $r_e$ value corresponds to equilibrium rotational constants of 0.3603 cm$^{-1}$
and 0.3599 cm$^{-1}$ for $^{161}$DyO and $^{163}$DyO, respectively, 
which are in reasonable agreement with the measured values of 0.3583 cm$^{-1}$
and 0.3579 cm$^{-1}$ derived from the measured rotational constants in Table \ref{tab:RotationalFit}. 
The computed harmonic frequencies and anharmonic constants take the values of 846.6 cm$^{-1}$ and 2.4 cm$^{-1}$ for $^{161}$DyO
and the values of 846.2 cm$^{-1}$ and 2.4 cm$^{-1}$ for $^{163}$DyO.
Thus, the fundamental vibrational frequencies take values of 841.9 cm$^{-1}$ and 841.4 cm$^{-1}$
for $^{161}$DyO and $^{163}$DyO, respectively. They agree closely with the measured values
of 842.4 cm$^{-1}$ and 841.9 cm$^{-1}$ in Table \ref{tab:RotationalFit}.
The X2CAMF-CCSD(T) electric dipole moment value of 4.58 Debye also shows close agreement
with the measured value of 4.507(26) Debye~\cite{linton_stark_1992}.

\begin{table}[tb]
\setlength{\tabcolsep}{12pt}
		\caption{The $h$ and $eQq_0$ values in units of \wn for the X8 state
        computed at the X2CAMF-CCSD(T) bond length of 1.793 \AA. The X2CAMF scheme has been used to treat relativistic effects. }
        \label{tab:eQqcomp}
\begin{tabular}{ccc}
\hline \hline
 $h$     &   $^{161}$DyO & $^{163}$DyO  \\     
 \hline
HF       & $-0.0469$         & $0.0658$            \\
CCSD     & $-0.0491$         & $0.0688$           \\
CCSD(T)  & $-0.0493$         & $0.0690$            \\
Measured & $-0.05109(49)$     & $0.07143(34)$          \\
\hline \hline
 $eQq_0$ & $^{161}$DyO & $^{163}$DyO \\
 \hline
HF       &      $0.0198$         & $0.0209$           \\
CCSD     &      $0.0191$         & $0.0202$            \\
CCSD(T)  &     $0.0161$        & $0.0170$            \\
Measured &     $0.0096(35)$        & $0.0098(25)$            \\
\hline \hline
\end{tabular}
\end{table}

Now we compare the computed $h$ and $eQq_0$ values to the measured ones. 
The strength of the magnetic hyperfine interaction and nuclear quadrupole interaction can provide a useful validation of the molecular structure calculations used to infer sensitivity to $CP$-violating physics, which also depends strongly on 
the distribution of the electron cloud in the vicinity of the Dy nucleus.
As summarized in Table \ref{tab:eQqcomp}, the X2CAMF-CCSD(T) calculations predict $h^{(161)} = -0.0493$~\wn, $h^{(163)} = 0.0690$~\wn for the electronic ground $X8$ states of DyO.
These computational values are in excellent agreement with the experimentally determined values
of -0.0511(4)~\wn and 0.0711(4)~\wn. 
A close inspection of the HF, CCSD, and CCSD(T) values reveals that the computational results are well convergenced with respect to treatments of electron correlation. The electron-correlation contributions defined as the difference between CC and HF values account for a small fraction of the total values. The close agreement between the CCSD(T) and CCSD values indicates that the remaining correlation contributions are small. 

The X2CAMF-CCSD(T) results of 0.016~\wn and 0.017~\wn for Dy $eQq_0$ values in $^{161}$DyO and $^{163}$DyO are qualitatively in agreement with the experimental values of 0.0096(35)~\wn and 0.0098(25)~\wn. 
We have performed CCSD(T) calculations using the X2C model potential (X2CMP) scheme
\cite{vanWullen05,Liu06,Knecht22,Wang25} as implemented 
in Ref.~\onlinecite{Wang25} as well as HF calculations using the four-component Dirac-Coulomb-Gaunt (4c-DCG) Hamiltonian
to investigate the remaining relativistic corrections. 
The X2CMP scheme covers multi-center relativistic two-electron contributions and is more accurate than the X2CAMF scheme
featuring more efficient one-center approximation for two-electron spin-orbit contributions. 
The 4c-DCG approach represents rigorous treatments of relativistic effects. 
The X2CMP-CCSD(T) $eQq_0$ values are larger than the X2CAMF-CCSD(T) ones by 0.0002~\wn. 
The 4c-DCG HF values are smaller than the X2CMP-HF ones by $-0.0002$~\wn. Therefore,
it is safe to conclude that the remaining relativistic corrections beyond the X2CAMF scheme are insignificant. 
On the other hand, there may be room for improvement on the treatments of electron correlation effects. As shown in Table \ref{tab:eQqcomp}, the CCSD correlation contributions to the $eQq_0$ values are unusually small. This can be tentatively attributed to accidental cancellation of the electron correlation contributions at the CCSD level. The non-iterative triples contributions from CCSD(T) appear to be significant and amount to around 20\% of the total value. It is worthwhile investigating contributions from high-level correlation contributions beyond CCSD(T) in future work. Further benchmarking and validation of quantum-chemical calculations with higher accuracy will also benefit from improved experimental determination of $eQq_0$ to the level of $\sim$10\% relative uncertainty or better.

\section{Conclusion}
In summary, we have measured hyperfine-resolved spectra of low-lying rotational levels in the (0,0) and (0,1) vibronic bands of the [17.1]7$\leftarrow$X8 electronic transition. The (0,1) band exhibits resolvable isotope shifts, which makes assignment of quantum states tractable. Main transitions with $\Delta F = \Delta J$ are identified by large transition strengths at low laser power, while satellite transitions with $\Delta F \neq \Delta J$ are observed at high laser power and enable direct measurements of hyperfine splittings within a rovibronic state using combination differences. We perform a global analysis of 195 observed transitions, resolving magnetic and electric hyperfine parameters, $h$ and $eQq_0$, in all three measured vibronic states. The hyperfine parameters determined in this way compare favorably to high-level \textit{ab initio} results, thus serving as a benchmark of the molecular structure calculations required to interpret future precision measurements with DyO. We also rationalize the measured ground-state values on the basis of a simple model of the molecular orbitals, which illustrates the dominant electronic configuration in the X8 state.

This study lays the groundwork for future experiments investigating the excited [18.0]9, [18.5]9, and [19.0]9 states, which all appear promising for optical cycling. We are currently pursuing high-resolution spectroscopy of the [18.0]9$\leftarrow$X8 transition, including measurements of branching fractions and the radiative lifetime, determination of the hyperfine structure, and measurements of the electric dipole moment and magnetic $g$-factor. In addition to providing practical information for optical cycling and precision searches for \T-violating new physics, these measurements will provide high-precision data that can be used to further benchmark relativistic quantum chemical calculations.

\section*{Supporting Information}
The measured transition wavenumbers, associated assignments, and difference between the observed and calculated transition wavenumbers are given in Tables S1--S4. The data can be accessed as an online resource at Ref.~\cite{SupportingInfo}.

\section*{Acknowledgments}
We are grateful to Prof.\ Tim Steimle and Prof.\ Tom Varberg for valuable contributions to the experimental apparatus, and to Tim Steimle for insightful conversations about the interpretation of hyperfine constants. We thank Jason Mativi, Michael Taylor, and Dave Williams for critical technical support. The experimental work was supported by the National Science Foundation under Grant No. PHY-2513425. The computational work at Johns Hopkins University was supported by the National Science Foundation under Grant No. PHY-2309253.

\bibliography{DyO_library}

\section*{TOC Image}
\begin{figure}[b]
    \centering 
    \includegraphics[width=0.9\linewidth]{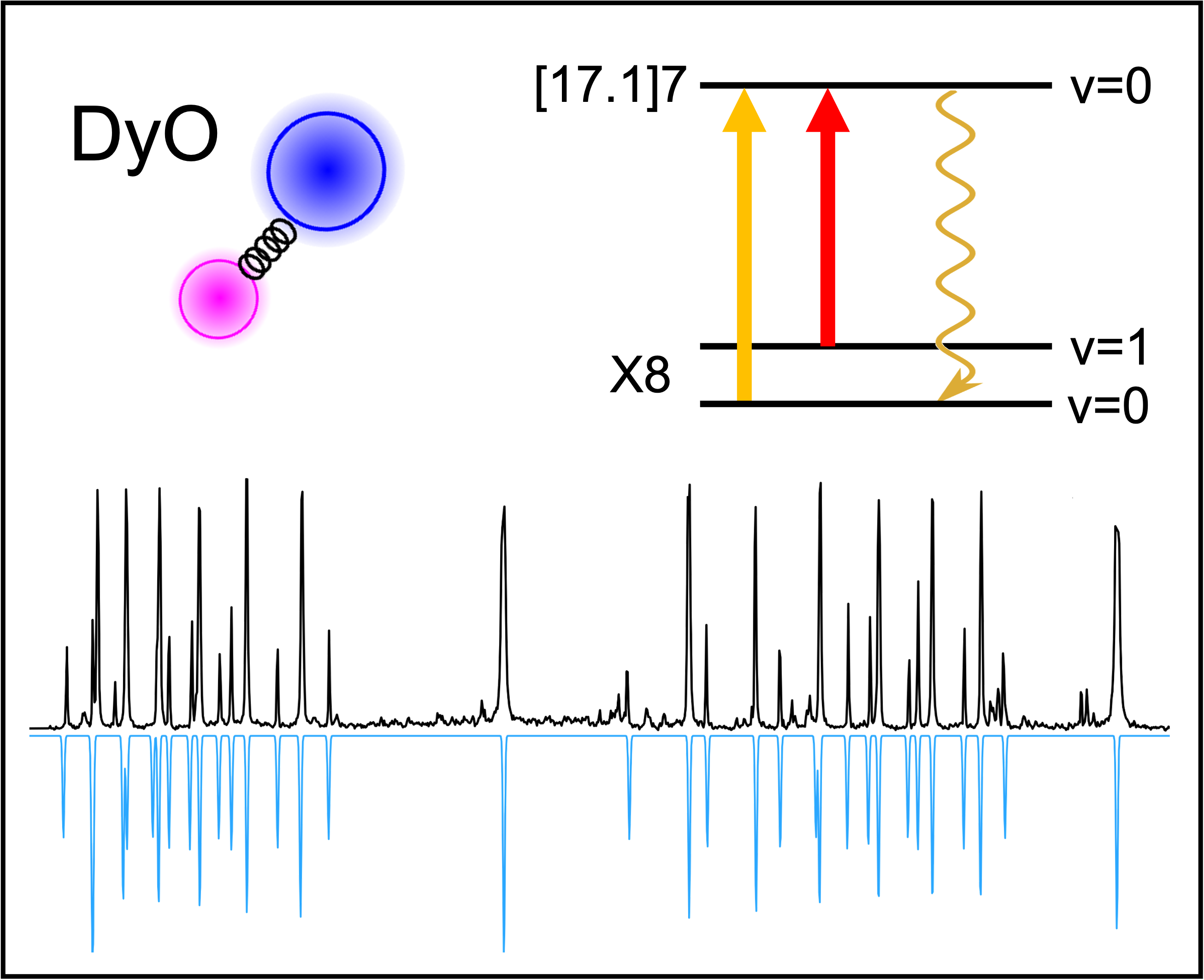}
    \label{fig:TOCImage}
\end{figure}

\end{document}